\documentstyle[12pt,epsfig]{article}

\def\thefootnote{\fnsymbol{footnote}}

\begin{document}

\newcommand{\beq}{\begin{equation}}
\newcommand{\eeq}{\end{equation}}
\newcommand{\beqn}{\begin{eqnarray}}
\newcommand{\eeqn}{\end{eqnarray}}
\newcommand{\ra}{\rightarrow}
\newcommand{\lra}{\leftrightarrow}
\newcommand{\tr}{{\rm Tr}}

\newcommand{\su}{$ SU(2) \times U(1)\,$}
\newcommand{\gag}{$\gamma \gamma$ }
\newcommand{\see}{\sqrt{s_{ee}}}
\newcommand{\sgg}{\sqrt{s_{\gamma\gamma}}}

\newcommand{\dkg}{\Delta \kappa_{\gamma}}
\newcommand{\dkz}{\Delta \kappa_{Z}}
\newcommand{\dz}{\delta_{Z}}
\newcommand{\dgz}{\Delta g^{1}_{Z}}
\newcommand{\la}{\lambda}
\newcommand{\lag}{\lambda_{\gamma}}
\newcommand{\laz}{\lambda_{Z}}
\newcommand{\lambdae}{\lambda_{e}}
\newcommand{\lnl}{L_{9L}}
\newcommand{\lnr}{L_{9R}}
\newcommand{\lt}{L_{10}}
\newcommand{\lu}{L_{1}}
\newcommand{\ld}{L_{2}}
\newcommand{\bbbar}{b\overline{b}}
\newcommand{\ccbar}{c\overline{c}}

\newcommand{\np}{{\em Nucl.\ Phys.\ }}
\newcommand{\pl}{{\em Phys.\ Lett.\ }}
\newcommand{\pr}{{\em Phys.\ Rev.\ }}
\newcommand{\prl}{{\em Phys.\ Rev.\,Lett.\ }}
\newcommand{\prep}{{\em Phys.\ Rep.\ }}
\newcommand{\zp}{{\em Z.\ Phys.\ }}
\newcommand{\sovjnp}{{\em Sov.\ J.\ Nucl.\ Phys.\ }}
\newcommand{\nuclinst}{{\em Nucl.\ Instrum.\ Meth.\ }}
\newcommand{\annp}{{\em Ann.\ Phys.\ }}
\newcommand{\intjmp}{{\em Int.\ J.\ of Mod.\  Phys.\ }}

\newcommand{\eps}{\epsilon}
\newcommand{\mhmh}{M_{H}^2}
\newcommand{\mh}{M_{H}}
\newcommand{\Mh}{$M_{H}\;$}
\newcommand{\Mhe}{M_{H}}
\newcommand{\mt}{M_{t}}
\newcommand{\sh}{$\hat{s}\;$}
\newcommand{\she}{\hat{s}}
\newcommand{\mw}{M_{W}}
\newcommand{\mww}{M_{W}^{2}}
\newcommand{\mwmw}{M_{W}^{2}}
\newcommand{\mz}{M_{Z}}
\newcommand{\mzz}{M_{Z}^{2}}
\newcommand{\cl}{{{\cal L}}}
\newcommand{\cd}{{{\cal D}}}
\newcommand{\cv}{{{\cal V}}}
\newcommand{\lgg}{\lambda_1\lambda_2}
\newcommand{\lww}{\lambda_3\lambda_4}

\newcommand{\ppin}{ P^+_{12}}
\newcommand{\pmin}{ P^-_{12}}
\newcommand{\ppout}{ P^+_{34}}
\newcommand{\pmout}{ P^-_{34}}
\newcommand{\sinsq}{\sin^2\theta}
\newcommand{\cossq}{\cos^2\theta}
\newcommand{\yt}{y_\theta}
\newcommand{\hppll}{++;00}
\newcommand{\hpmll}{+-;00}
\newcommand{\hpplt}{++;\lambda_30}
\newcommand{\hpmlt}{+-;\lambda_30}
\newcommand{\hpptt}{++;\lambda_3\lambda_4}
\newcommand{\hpmtt}{+-;\lambda_3\lambda_4}
\newcommand{\dk}{\Delta\kappa}
\newcommand{\klam}{\Delta\kappa \lambda_\gamma }
\newcommand{\kac}{\Delta\kappa^2 }
\newcommand{\lac}{\lambda_\gamma^2 }

\def\slashc{c\kern -.400em {/}}
\def\slashL{L\kern -.450em {/}}
\def\slashcl{\cl\kern -.600em {/}}
\def\gam{\gamma \gamma}
\def\ggzz{$\gamma \gamma \ra ZZ \;$}
\def\gag{$\gamma \gamma\;$}
\def\ggww{$\gamma \gamma \ra W^+ W^-\;$}
\def\ggwwe{\gamma \gamma \ra W^+ W^-}
\def\W{{\bf W}}
\def\B{{\bf B}}
\def\noi{\noindent}
\def\sm{${\cal{S}} {\cal{M}}\;$}
\def\smx{{\cal{S}} {\cal{M}}\;}
\def\nph{${\cal{N}} {\cal{P}}\;$}
\def\ssb{${\cal{S}} {\cal{B}}\;$}
\def\cviol{${\cal{C}}\;$}
\def\pviol{${\cal{P}}\;$}
\def\cpviol{${\cal{C}} {\cal{P}}\;$}
\newcommand{\epm}{$e^{+} e^{-}\;$}

\newcommand{\epemt}{$e^{+} e^{-}\;$}
\newcommand{\epem}{e^{+} e^{-}\;}
\newcommand{\gamgamt}{$\gamma \gamma \;$}
\newcommand{\gamgam}{\gamma \gamma \;}
\newcommand{\egamt}{$e \gamma \;$}
\newcommand{\egam}{e \gamma \;}
\newcommand{\ggwwt}{$\gamma \gamma \ra W^+ W^- \;$}
\newcommand{\ggwwht}{$\gamma \gamma \ra W^+ W^- H \;$}
\newcommand{\ggwwh}{\gamma \gamma \ra W^+ W^- H \;}

\newcommand{\ptu}{p_{1\bot}}
\newcommand{\vecptu}{\vec{p}_{1\bot}}
\newcommand{\ptd}{p_{2\bot}}
\newcommand{\vecptd}{\vec{p}_{2\bot}}
\newcommand{\ie}{{\em i.e.}}
%************

%\setcounter{secnumdepth}{2}
%\addcontentsline{toc}{section}{APPENDIX}

\def\sss{\scriptscriptstyle}
\def\lft{{\sss L}}
\def\rht{{\sss R}}
\def\subs{{\sss S}}
\def\subt{{\sss T}}

\newcommand{\bibit}{\nineit}
\newcommand{\bibbf}{\ninebf}
%\renewenvironment{thebibliography}[1]
% { \elevenrm
%   \begin{list}{\arabic{enumi}.}
%    {\usecounter{enumi} \setlength{\parsep}{0pt}
%     \setlength{\itemsep}{3pt} \settowidth{\labelwidth}{#1.}
%     \sloppy
%    }}{\end{list}}

\begin{titlepage}

%%%%%%%%%%%%%%%%%%%%%%%%%%%%%%%%%%%%%%%%%%
%%%%%%  THIS FILE SHOULD BE TYPESET WITH LATEX      %%%%%%%
%%%%%%       run twice
%%%%%%%%%%%%%%%%%%%%%%%%%%%%%%%%%%%%%%%%%%

\newcommand{\norm}[1]{{\protect\normalsize{#1}}}
\newcommand{\LAP}
{{\small E}\norm{N}{\large S}{\Large L}{\large A}\norm{P}{\small P}}
\newcommand{\LAPP}
{{\scriptsize E}{\small N}{\norm S}{\large L}{\norm A}{\small P}{\scriptsize
P}}

\renewcommand{\thefootnote}{\fnsymbol{footnote}}
\newpage
\pagestyle{empty}
\setcounter{page}{0}
%%%%%%%%%%%%%%%%%%%%%%%%%%%%%%%%%%
%%%%%%%%%%%  ENTETE ENSLAPP  %%%%%%%%%%%%
%%%%%%%%%%%%%%%%%%%%%%%%%%%%%%%%%%

%\input epsf
\begin{minipage}{4.9cm}
\begin{center}
{\bf  G{\sc\bf roupe} d'A{\sc\bf nnecy}\\ \ \\
Laboratoire d'Annecy-le-Vieux de Physique des Particules}
\end{center}
\end{minipage}
\hfill
%%%\raisebox{-1.2cm}{\epsfbox{enslapp.ps}}
\hfill
\begin{minipage}{4.2cm}
\begin{center}
{\bf G{\sc\bf roupe} de L{\sc\bf yon}\\ \ \\
Ecole Normale Sup\'erieure de Lyon}
\end{center}
\end{minipage}

\begin{center}
\rule{14cm}{.42mm}
\end{center}
%%%%%%%%%%%%%%%%%%%%%%%%%%%%%%%%%%
%%%%%%%%%%%  ENTETE ENSLAPP FIN %%%%%%%%%%
%%%%%%%%%%%%%%%%%%%%%%%%%%%%%%%%%%

\relax
%%%%%%%%%

\begin{center}

{ \Large \bf Symmetry breaking and electroweak physics at
 Photon Linear Colliders}

\vspace*{1.cm}

\begin{tabular}[t]{c}

{\bf G.~B\'elanger}\\
\\
%\vspace*{0.5cm}

{\it  Laboratoire de Physique Th\'eorique}
EN{\large S}{\Large L}{\large A}PP
\footnote{URA 14-36 du CNRS, associ\'ee \`a l'E.N.S de Lyon et \`a
l'Universit\'e de Savoie.}\\
{\it Chemin de Bellevue, B.P. 110, F-74941 Annecy-le-Vieux, Cedex, France.}

\end{tabular}
\end{center}
\vspace*{\fill}

\centerline{ {\bf Abstract} }
\baselineskip=14pt
\noindent
%%%%%%%%%%%%%%%%%%%%%%%%%%%%%%%%%%%%%%%%%%%%%%%%%%%%%%%%%%%%%%%%%%%%%
 {\small
The physics potential of a high-energy photon collider
is reviewed. The emphasis is put on aspects related to the symmetry
breaking sector, including Higgs searches and  production of
longitudinal vector bosons.
}

\vspace*{\fill}

%\hrulefill\ $\; \; \; \; \; \; \; \; \; \; \; \; \; \; \; \; \; \; \; \;\;$
%\hspace*{3.5cm}\\

\vspace*{0.1cm}
\rightline{ENSLAPP-A-527/95}
\rightline{arch-ive/9508218}
\rightline{June 1995}

\noindent
{\footnotesize $\;^{\S}$ URA 14-36 du CNRS, associ\'ee \`a l'E.N.S de Lyon et
\`a
l'Universit\'e de Savoie.}\\
\noindent
{\footnotesize Talk presented at the Photon'95 Conference,
 Sheffield, England, April 8-15, 1995.}
\vspace*{\fill}
\end{titlepage}
\baselineskip=18pt

\baselineskip=14pt

%*************
\setcounter{section}{1}
\setcounter{subsection}{0}
\setcounter{equation}{0}
\def\thesubsection {\thesection.\arabic{subsection}}
\def\theequation{\thesection.\arabic{equation}}

\setcounter{equation}{0}
\def\thequation{\thesection.\arabic{equation}}

\setcounter{section}{0}
\setcounter{subsection}{0}

\section{Introduction}

The option of
building  a photon-photon interaction region
at an \epm linear collider is now taken seriously under consideration.
Based on
the idea of using laser-induced backscattered photons for
inducing high-energy photon collisions,
 a \gag collider (PLC)
gives rise to new physics opportunity\cite{GinzburgNIM}.
The issues concerning  electroweak physics
 will be summarized in this talk\cite{Paris}.

Since the symmetry-breaking mechanism remains the last open
question in the standard model,
 an important part of the planning at any future
collider must be devoted  to that.
 This obviously includes   searches for the Higgs particle
and the determination of its properties.
One unique opportunity for \gag colliders in that respect is the
direct measurement of the $H\gam$ coupling.
Should the Higgs searches remain fruitless, the
 study
of the longitudinal $W$ sector would  give
a handle on the symmetry breaking mechanism. Photon
 colliders, being essentially a $W$ pair
factory, could make a useful
 contribution in that respect.

Before going into the heart of the subject
and to give a first idea of the possibilities of
\gag colliders,  I will present the main characteristics of
high-energy \gag collisions:
\begin{itemize}
\item
Any elementary charged particle, phase-space allowing,
 can be produced in \gag collisions with a model-independent
predictable cross-section.

\item
 \gag gives access to the  $J_Z=0$ channel, which is chirality suppressed
in \epm. To test the electroweak  symmetry breaking (ESB) mechanism,
this means producing the Higgs as  a resonance.

\item
\gag collisions feature  very large cross-sections, which are
always larger than in \epm for the
same energy and luminosity.
\end{itemize}

But all is not so bright, for \gag colliders also have  shortcomings.
First,
the  Higgs resonance cannot be so prominent as the Z at LEP since the coupling
of
neutral scalars to two photons only occurs at the loop
level and is suppressed by a factor $\alpha$.
Second, \gag does not have the same energy as \epm
(the maximum energy varies between 80-90\%) and the useful luminosity
can be smaller than in \epm. The latter is true especially if
one uses beams optimized for \epm coliders rather then designed
specifically for \gag\cite{Telnovshef}.
Finally, the photon collider is not monochromatic,
although, as was discussed by Telnov,
 one can tune the parameters of the
laser  such as to have a nearly monochromatic
spectrum near the maximum energy. This is done at the expense
of a drop in luminosity.

Certainly,  one has a great
flexibility in choosing the energy, the spectrum and
the polarization of the beams. The
choice of spectrum will be dictated by the physics one is interested in.
For example, a ``peaked" spectrum where much the luminosity is
 concentrated over a narrow
energy band would be most appropriate to study
a resonance.  A ``broad" spectrum, one
with sizeable luminosity over a wide
energy range, would correspond to a multi-purpose
machine useful for many processes\cite{Paris}.
Whatever the spectrum used, a precise knowledge of it is
essential to be able to do precision measurements.
More efforts in that direction are needed since
recent studies have shown that going beyond the ideal spectrum of
Ginzburg et al.\cite{GinzburgNIM}
could significantly affect the region where the photons carry only a
small fraction of the initial beam energy.
This region is particularly sensitive to
 multiple scattering and nonlinear effects\cite{Chenshef}.
For lack of a more realistic  spectrum, most of the results  presented
here use the ideal one.

\section{Typical  electroweak cross-sections}
\renewcommand{\thequation}{\thesection-\arabic{equation}}
\setcounter{equation}{0}

A comparison of a few characteristic
cross-sections in \gag with the corresponding same final-state
processes at \epm clearly show the advantage of the former.
Indeed,  independent of the spin of the particle
and at the same centre-of-mass
energy, \gag-initiated processes are, at high enough energy, about an order of
magnitude larger than the corresponding \epm reactions
(Fig.~1).
 Even the reaction
$\gamma \gamma \ra ZZ$ \cite{Jikiazz,DicusKao},
which is purely a loop effect, rapidly overtakes the
corresponding tree-level \epm process. This is due
to the rescattering effect $\gamma \gamma \ra W^+W^- \ra ZZ$.
Vector-boson production
dominates in \gag collisions due to the t-channel spin-1 exchange. Most
prominent is the $W$ pair cross-section, which very quickly reaches a plateau
of almost $90$pb.
This process is so important that it triggers a host of higher order
processes like
triple vector production ($\approx 1$pb), 4 vector production
($\approx 100 fb$) or  $H$ production via $WWH$ \cite{nousgg3v}.

\begin{figure*}[htb]

\caption{\label{allgamegam}
{\em Typical sizes of non-hadronic \gag, $e\gamma$ and
\epm processes.
The subscripts in Higgs(top)  processes refer to the mass of the
Higgs(top).  }}
\begin{center}
\vspace*{.5cm}
%%%\mbox{\epsfxsize=12.cm\epsfysize=13.2cm\epsffile{ggpro.eps}}
\vspace*{-1.cm}
\end{center}
\end{figure*}

Large cross-sections are great, but more
 is required to do interesting physics.
 A drawback of
\gag processes, especially as regards ESB
tests is the small fraction of longitudinal
vs transverse $W$'s.
While there is a large sample of
$W_LW_L$ (in fact more then 5 times than in $e^+e^-$),
the extraction of these longitudinals from the transverse background is
 an
arduous task. Since most transverse W's are produced quite forward, imposing
angular cuts improves the situation significantly (see
Fig.~2). Nevertheless, the ratio LL/TT remains higher in \epm.
The problem just alluded to is in fact characteristic
of all processes that will be discussed for ESB tests in
\gag: the extraction of a signal, usually in the longitudinal sector, from
a large transverse background.

\begin{figure*}[hbt]
\vspace*{-.5cm}
\begin{center}
\caption{\label{eeggwwfig2}{\em Comparing the total $WW$ cross-sections
and the longitudinal $W_L W_L$ in \epm {\it vs} \gag.
}}
%%%\mbox{\epsfxsize=13cm\epsfysize=8.cm\epsffile{eeggwwfig2.eps}}
\vspace*{-1.cm}
\end{center}
\end{figure*}

 \section{Electroweak symmetry breaking}

The various options for symmetry breaking can be divided into two
classes:
 light Higgs ($m_H\leq 800 GeV$)
or
 no Higgs (for the discussion here this is equivalent to a
 heavy Higgs).
 The implications for electroweak physics differ markedly
 according to the option one is willing to consider.
 A light Higgs probably means the existence of supersymmetry    unless
 one is ready to give up the naturality argument raised to avoid the
 large fine-tuning necessary for the elementary
 scalar to remain light to all orders.
 In this option the physics issues at a  collider, in particular
 the PLC, would be the
 search for the Higgs and measurement of  its properties,
 the search for other supersymmetric particles
 and determination of  their properties, and   precision measurements
 in order to see indirect effects of new physics. An example of the latter is
the
  measurement of the  trilinear couplings, $WW\gamma$.
  These topics can be covered at moderate
  energies linear colliders ($\see=300-500 GeV$)

  If the light Higgs does not exist, then  ESB is triggered by strong forces,
the scale being set by
  $\Lambda=4\pi v\approx 1$~TeV. Although the details of the model are
not known
 there must be new physics at this scale (e.g. technicolour,
strongly-interacting particles).
In particular, this new physics would show up in $W$ self-interactions
or in $W_LW_L$ scattering.
The connection between heavy Higgs and longitudinal W's is best
established via the process
$W_L W_L\ra W_L W_L$.
In the SM the Higgs is introduced to cure the bad high-energy behaviour
of this amplitude; nonetheless
$W_LW_L$ interactions become strong if $m_H\approx .8-1$~TeV.
The physics of the heavy Higgs is most relevant for TeV linear colliders.

Since no model has gained a consensus to
describe the strongly-interacting electroweak sector, one must
strive for  a model-independent
description of this sector.
One approach, which is valid up to some
scale $\Lambda$, uses an  effective  chiral Lagrangian.
Assuming a custodial $SU(2)$ symmetry
to ensure that the parameter $\rho\approx 1$,
new physics in the weak boson sector  is described by
nonrenormalizable terms suppressed by powers of
$1/\Lambda$,
\beqn
{\cal L}={\cal L}_{SM} (no Higgs)+\sum \frac{1}{\Lambda^n} {\cal L}_n
\eeqn
\noindent
The leading order chiral Lagrangian, ${\cal L}_2$,
contains only the mass terms for the vector bosons
while the Next-to-Leading order,
  ${\cal L}_4$, contains the self-interactions. This
includes trilinear or quartic interactions of massive vector bosons
and at most one photon\cite{Hawai}. Self-interactions including two photons
only appear at higher order, ${\cal L}_6$.

The effective Lagrangian formalism would break down if the
scale at which the experiment is performed is sufficient to produce
new resonances. These must be explicitly incorporated,
the cases
%To  incorporate the resonances
%in a model independent fashion  can be done once the choice
of either  a scalar, vector or tensor resonance will be considered.
The scalar one  ($\sigma$-like) is  representative of
a heavy Higgs while the vector one ($\rho$-like)
occurs in technicolour.
%more generally in the BESS model\cite{BESS}.

To
cover all standard and non-standard manifestations of symmetry breaking,
the strategy at the  future collider must include: tests of $W$
self-interactions and of longitudinal
vector-boson scattering, the search for the Higgs in the whole
range of possible masses, as well as searches for other heavy resonances.
All these aspects can be tackled at a photon collider,
as will be described in the rest of this talk.

 \section{\ggww and W self-interactions}

The importance of this process cannot be over-emphasized considering the
large cross-section involved.
Although the bulk of the reaction is due to the gauge transverse sector,
the fact that there are so many $W$'s around
makes this reaction the ideal place
 to conduct precision tests of the electromagnetic
couplings of the $W$.
In the effective chiral Lagrangian description of Higgsless models,
the anomalous trilinear couplings invoke two C and P conserving operators at
Next-to-Leading order,
$L_{9L}$ and $L_{9R}$,  and one C and P violating
 operator $L_{C}$\cite{Hawai}.
The latter affects only $ZWW$ and  $\gamma ZWW$ interactions while only
the combination  $L_{9L}+L_{9R}$ contribute to $\gamma WW$.
The $L_i$ operators are expected to be ${\cal O} (1)$.

There is an extensive literature
\cite{Paris}  on the effect of
anomalous couplings in various experiments.
Comparisons of different analyses have shown that a
$500$~GeV linear collider with  a luminosity
of $10 fb^{-1}$ does significantly better then the LHC for trilinear
couplings. Furthermore, the
limits that can be obtained in \gag at this energy,
$
|L_{9L}+L_{9R}|< 10
$
represent a 50\% improvement over \epm\cite{Hawai}. This is shown in
 Fig.~3.
However, this result is not sufficient  to reach the level
where one expects new physics to set in.
 In \epm it was shown  recently\cite{steve,barklow} that meaningful
limits could be obtained with a
luminosity ${\cal L}=50-80 fb^{-1}$.
It remains to be seen if the same
can be done at a \gag collider. For that one needs to
generate the four-fermion final state from the decay of the
$W$'s, while keeping the full spin-correlation.

\begin{figure*}[htb]
\caption{\label{gg19}{\em Comparison between the expected bounds on the
two-parameter space $(L_{9L},L_{9R})$ at the NLC500, LHC
and LEP2.
We also show (``bars") the limits from a single parameter fit.}}
\begin{center}
\vspace*{-1.cm}
%%%\mbox{\epsfxsize=12.cm\epsfysize=13.8cm\epsffile{l995.eps}}
\vspace*{-1.cm}
\end{center}
\end{figure*}

A \gag collider can do more than precision tests on $\gamma WW$.
It is  also sensitive to $ZWW$ couplings through processes with three
particles in the final state, for example
$\gam \ra WWZ$\cite{nousggwwz}, $e\gamma \ra e WW$ and $e\gamma\ra \nu W Z$.
 The latter
 is very sensitive to  the  operator
$L_{C}$\cite{Dawsong5}. The limits obtained, $L_{C}<25$
 at 500 GeV are comparable to the ones from
$e^+e^-\ra W^+W^-$\cite{BMT}. Furthermore, the sensitivity
increases rapidly with energy.

\subsection{Effect of radiative corrections}

When doing precision tests
one must worry about the effect of radiative corrections that could
mimic those of the new couplings.
Recently the complete one-loop SM corrections for helicity amplitudes
for \ggww were calculated\cite{Dennerradcor}. It turns out that the
radiative corrections for this
process are theoretically clean
 due to the absence of most
universal leading corrections.
The running of $\alpha$ is irrelevant since we are dealing with  on-shell
photons,
all uncertainties due to
$\log(m_q^2)$ terms in small masses disappear, and
there are no large log corrections associated with colinear
photons except at very high energies. Furthermore, the corrections are not very
sensitive to either $\mt$  or $\mh$ except near the resonance.
Although
some helicity amplitudes receive huge corrections,
they  are precisely the ones
that contribute very little to the total cross-section.
Typically  radiative corrections between $1-10\%$ at
$\see=500$~GeV are obtained, and they  tend to increase with
energy ($\approx 20\%$ at $1$~TeV).
In any case, the
 inclusion of radiative corrections are not expected to  change much
the previously obtained  results on measurements of trilinear couplings.

Considering the large
numbers of $W$'s available, there are other interesting questions
 that can be studied in \gag which I
have not addressed here. Among them are the possibility of
 direct tests of quartic couplings involving photons\cite{nousggvv},
 CP tests in $W$ decay
and measurement of the $Wtb$ coupling which could also give some clues
about symmetry breaking.

\section{Higgs searches}

One of the most attractive motivations for doing physics with very
energetic
photon beams is the unique capability of this mode for producing a
scalar particle, such as the Higgs, as a resonance.
I have already mentioned that the coupling of
the Higgs to two photons occurs only at the loop level.
It should be emphasized that a precision measurement of the
$H\gam$ coupling is an indirect way of revealing all massive charged
particles that could be present in an extension of the standard model.
These heavy quanta would not decouple and would
contribute to the production rate in \gag.

While many processes are sensitive to the presence of the Higgs
(see Table 1), the prime interest of the photon mode lies in the
Intermediate Mass region
\footnote{\gag is also very useful in the mass range below 90 GeV,
a case can be made for building a low-energy dedicated \gag collider
in the event of a discovery of the Higgs at LEP2.}.
For such a Higgs, the main decay mode is into $\bbbar$.
Although a search is feasible at LHC, it will be a difficult and long task to
extract
a signal in this case.
For heavier Higgs masses the resonance can be seen in the $WW$
\cite{Ginzburgshef} or $ZZ$ channel.
However, the usefulness of these modes is tamed by the presence of
large
backgrounds from transverse vector bosons.  Ultimately,
 for a Higgs above 400 GeV, and regardless
of the energy available for the PLC, one would have to resort
to other channels such as associated Higgs production
or $WW$ fusion (via the process $\gam\ra WWWW$).

\baselineskip=12pt
\begin{table*}[htb]
\caption{\label{Higgsmass}
{\em Processses for Higgs searches at PLC and other colliders}}
\vspace*{0.3cm}
\centering
\begin{tabular}{|l|l|l|l|}
\hline
Mass&Collider& PLC&$\sqrt{s_{ee}}$
\\
\hline
&&&\\
$\mh<65 GeV$& LEP& Ruled out&-----\\
$65 GeV<\mh<90GeV$&LEP2&$\gam\ra H\ra\bbbar$&$.1-.5$~TeV\\
$90 GeV<\mh<140GeV$&NLC & $\gam\ra H\ra\bbbar$&$.2-.5$~TeV \\
$140 GeV<\mh<200GeV$&LHC &$\gam\ra H\ra WW$&$.5$~TeV \\
$200 GeV<\mh<400GeV$&LHC &$\gam\ra H\ra ZZ$&$1$~TeV \\
$400 GeV<\mh<700GeV$&LHC  & $\gam\ra WWWW$&$2$~TeV\\
\hline
\end{tabular}
\end{table*}
\baselineskip=14pt

\subsection{Intermediate Mass Higgs}

For  the IMH, \gag can contribute in the discovery mode
or  perform precision measurements of its properties.
A crucial point relates to the choice of the spectrum
and polarization used. Since the Higgs is produced only
  in the
$J_z=0$ channel, polarization plays a crucial role in
enhancing the signal over background.
Assuming the Higgs has been found and its mass
measured, one could
tune the energy of the collider and the parameters of the laser
such that the peak of luminosity lies precisely at $\sqrt{s_{\gam}}=\mh$.
This is obviously the preferred way to operate when measuring
  $H\gam$,
though  one has to realise that good
luminosity is required. Early estimates
for a 500 GeV collider and a luminosity of ${\cal L}=20 fb^{-1}$
give a 10\% precision on the width,\cite{Borden} the effect of background from
one
radiated gluon is discussed by Khoze \cite{Khozeshef}.
One disadvantage of operating in that mode is that the PLC would be run
at energies much below the nominal \epm energy,
precluding the study of interesting processes such as the $W$ pair
production
and other $W$ reactions that could occur at higher
energy. Furthermore, this low-energy narrow-band scheme could
render kinematically inaccessible some of the particles
that would only be probed indirectly in $H\gam$,
not to mention that the \gag mode, when operated in the full range of
energy, can access scalar particles that would be kinematically out of
reach in the \epm mode.

If one would have two interaction regions,
one devoted to \gag the other to \epm,
and if the Higgs has not been found elsewhere,
\gag could be used  to search for the Higgs. The
method
%One method consist
%in varying the parameters of the laser, for a fixed \epm energy,
%so as to do a scan over the interesting mass range. Another simpler
that allows for simultaneous studies
of processes at high energy consists of  running
 the PLC using a ``broad" spectrum
so that one would have reasonable luminosity over a range of energies.

The main problem in the Higgs searches, whatever the scheme used,
lies in the large background. The prominent one comes from
 direct QED $\gam \ra q \bar q$ production
where $q=b$ or other light quark flavours, in particular charm.
However this background can be dealt with since
the bulk of the cross-section is in the forward direction, so that a modest
angular cut  could efficiently  suppressed  this background
and would almost totally eliminate its
$J_Z=0$ contribution.
Therefore
 a spectrum with a predominantly $J_Z=0$
component would both enhanced the signal and reduce the
background.

When the PLC is run in the ``broad" spectrum mode there are other
more important backgrounds that have to be taken into
account\cite{Halzen,Higgsreson}.
They arise from the hadronic structure of the photon
which can resolve into a gluon or a quark with some spectator jets
left over. One then has to worry about $q \bar q$ production through $\gamma g$
as well as a host of 1-resolved and 2-resolved process.
These backgrounds dominate the signal. However,
 since  most of the resolved events are very boosted,
 judicious cuts can reduce it to a manageable
level. It was shown, using the ideal spectrum,
that at $500$~GeV with ${\cal L}=10 fb^{-1}$,
 one could obtain a good signal  for $M_H=110-140$~GeV
\cite{Higgsreson}. Furthermore,
the situation improves for a collider of lesser energy, due to
the reduced resolved background. For example,
at $350$~GeV a signal is easily extracted for the whole IMH range.
Of course this assumes  the
 ideal spectrum. However, for the masses
considered, the signal falls in the region where the spectrum is most
severely affected by effects of multiple scattering and nonlinear
effects. These questions should be reassessed taking these effects into
account as the conclusions could differ drastically.

There have been suggestions to determine directly
the parity of the Higgs using
linear polarizations of the photon\cite{Gunionparity}. Since the degree of
linear
polarization is never very large
($<30\%$), this always requires large luminosities,
${\cal L}=100 fb^{-1}$.
%Furthermore, the parity
%of the scalar will, in a large degree, be inferred from its rate of
%production in the \epm.

%To summarize, the best strategy if we will have two interaction
%5regions at a linear collider, with
%one devoted to
%\gag physics, would be to
% search in both
%modes with a later (long) run dedicated to precision measurements
%with $s_{\gam}\approx \mh$.

\subsection{Associated  production}

For the IMH, it will be hard to unravel a peak formation
if the collider energy is greater than $500$~GeV.
As will be discussed in the next section, the resonance will remain
hidden
for heavier Higgs ($\mh\ge 400$~GeV)
 even if one uses the most favourable channel, $ZZ$.
Fortunately, other efficient mechanisms for Higgs production are
available, in particular the
radiation of a Higgs from a $W$ pair.
This is to be expected since the cross section for $W$
 pair production is
so large and the Higgs couples preferentially to the weak bosons.
In fact, at
 1 TeV,  before folding with the luminosity
spectrum, the $e\gamma$, \gag and Bjorken process are comparable
for all Higgs masses\cite{nousgg3v}.
However, the \gag production mode is suppressed
 when including a more realistic photon luminosity
spectrum. Still,
at $1$~TeV one obtains a measurable cross-section ($\sigma >3fb$)
for $\mh<400 GeV$.

 \subsection{\boldmath{ $\gam \ra ZZ$}}

At first this reaction  was believed to
provide a background-free environment
for either Higgs production or non-standard physics
signals in $Z_LZ_L$ since it is purely a loop process
in the \sm.
The first full calculation by Jikia\cite{Jikiazz}
 of the one-loop process \ggzz within the \sm
dampened this enthusiasm since it
 turned out that, once again, the
transverse modes are overly dominant, especially at high energy.
 This is due essentially to the $W$
loops, the $WW$ produced in \gag rescatter
into $ZZ$.
At
 $\sqrt{s_{ee}}=400$~GeV,
the Higgs resonance is clearly evident over the $TT$ continuum all the way
up to the kinematic limit. With $\sqrt{s_{ee}}=500$~GeV,
 it already becomes difficult to extract a Higgs with $M_H\sim350$~GeV.
\cite{Jikiazz}
To obtain these results, Jikia used a predominantly $J_Z=0$
spectrum that is peaked towards the maximum $\hat{s}_{\gam}$,
this is not the optimum choice.
With  a broader spectrum featuring
a dominant  $J_Z=0$ for small $M_{ZZ}$,
one could still see a peak in the $M_{ZZ}$
invariant mass for Higgs masses up to $400$~GeV at a $1$~TeV \epm machine
\cite{DicusKao}.
{}From the perspective of observing the Higgs resonance beyond TeV
\epm energies, the situation
becomes totally hopeless as the
transverse $ZZ$ are awesome\cite{Jikiazz}.

%As for the signature, the cleanest signal that
%allows an invariant $ZZ$ mass reconstruction
%is when at least one of the $Z$ decays leptonically (one should only consider
%$e$ and $\mu$) while the other decays visibly even though
%5this corresponds
%to a small combined branching fraction of about $10\%$ of
% all $ZZ$ events\cite{Jikiazz}.

\section{Strongly-interacting electroweak sector (SEWS)}

If the Higgs is not found at LHC, or in the sub-TeV version of
NLC, we will be in the realm of the SEWS.
 This sector would be probed most efficiently
at TeV energies through the reaction
 $V_L V_L\ra V_L V_L $ ($V=W$ or $Z$).
In this channel one  would either search  for a resonance
or, if the energy is not sufficient, for new interactions such as the ones
described by the  effective chiral Lagrangian.

The $V$ pair-production processes could be regarded as the testing ground for
possible rescattering effects in $WW\ra VV$ that originate from the
symmetry breaking sector.
Unfortunately,
at high energies, {\it i.e.}, at high $VV$ invariant masses, where
the effect of the New Physics would be most evident, one has to fight
extremely hard against the background for transverse $W$ and $Z$.
Indeed,
recent analyses have shown that while it might be possible to see
effect of a tensor resonance, a scalar one as well as indirect effects
are
hopeless\cite{BergerBerkeley}.

%\begin{figure*}[hbt]
%\begin{center}
%\caption{\label{ggwwsplit}{\em A photon of helicity $\lambda$
%``splits" into a longitudinal $W$ that carries a fraction
%$y$ of its momentum and a spectator $W$ with helicity $\lambda'$.
%The longitudinal $W$ then takes part in the hard process.}}
% \vspace*{0cm}
% \mbox{\epsfxsize=15.5cm\epsfysize=6cm\epsffile{ggwwsplit.eps}}
% \vspace*{-.5cm}
%\end{center}
%\end{figure*}

One then has to resort to the only source of longitudinal vector
bosons, the ones taking part in the fusion process and contributing
to $WWWW$ or $WWZZ$ production. This process is the analog of $e^+e^-\ra
\nu\overline\nu W^+W^-$ and was originally believed to be
more favourable due to a presumed larger $W_L$ content in the
photon than in the electron.
While it is true that in the photon there is an additional
structure function corresponding to the spectator $W_L$,
the dominant contribution is from transverse
spectator $W$'s. The latter features basically the same structure
function as in the electron, except for an overall factor\cite{Paris}.
 One would therefore expect that
\gag should be comparable to \epm at the same energy and luminosity.
This conclusion was born out
by two independent exact calculations of this \sm process
\cite{Jikiawwww,Cheungwwww}.
The signal of a heavy Higgs is a significant increase
in the channels with at least three $W_L$.
To extract a signal requires tagging all four $W$'s, the spectator ones
being associated with
the low $p_T$ and the longitudinal ones with the central $W$'s.
The spectators are tagged with one hadronic and one leptonic decay while
the central ones go into four jets. The results of the analysis
showed that a 2~TeV PLC (${\cal L}=10 fb^{-1}$)
would give a good signal ($S/\sqrt{B}\approx 10$)
for a heavy Higgs-like scalar of 1TeV\cite{Cheungwwww}. This is  comparable to
the \epm process. However, the inclusion of the
spectrum has a dramatic effect and a linear collider of
$2$~TeV in the \epm center of mass with ${\cal L}=200fb^{-1}$
is needed to reach the same significance level.

Another interesting conclusion from these calculations is that a signal
for a Higgs of 400-700~GeV can easily be seen with
$\see=1.5$~TeV and ${\cal L}=200 fb^{-1}$.
The PLC can therefore cover the whole mass range for light or
heavy Higgs searches
provided a good choice of energy and spectrum is made,
although precision measurements are possible only for light Higgs
($\mh< 120 GeV$).

\section{Search for new particles: supersymmetry}

As the best motivated alternative to the standard model, one should
investigate the consequence of supersymmetic models.
Supersymmetry would provide a natural framework for
light Higgses.
The three neutral scalars of supersymmetric models, $h$,$H$,$A$
 (pseudoscalar)
could be produced as a resonance in \gag. This reaction would then extend
the reach in mass of \epm since in the latter $H$ and $A$ can only be
produced together and require
$\see>M_H+M_A$. At $\see=500$~GeV, using the $\bbbar$ mode, this gives
the following discovery region for the supersymmetric
scalars\cite{GunionHaber}:
$
110$~GeV$< \mh < 200$~GeV and  $100$~GeV$<M_A < 2 \mt$.
Recently, it was pointed out that this was true only if scalars decayed
primarily via \sm final states.
Otherwise the above limits require high luminosities ${\cal L}> 60
fb^{-1}$\cite{Gunionsusy}.

The \gag collider can search also for other supersymmetric particles
\cite{Murayama},the
production cross-sections being universal, were already shown.
Typically, one finds that \gag can have good cross-sections but offer
little advantage over the \epm mode, in part because of the
lower achievable energy. It is for selectron searches in $e\gamma\ra
\tilde{e}\tilde{\chi}$ that the laser scheme becomes extremely useful,
as the discovery of a selectron of
$m_{\tilde{e}}\approx\sqrt{s_{e\gamma}}$ is possible.

\section{Conclusion}

A \gag collider of energy ranging from .2 to 2~TeV
should prove to be a useful tool for probing the electroweak
symmetry-breaking sector through either Higgs
searches or $W$ physics. It is  unique in producing
a scalar on resonance and is complementary
to an  \epm collider in many  processes.

\noindent
{\bf {\large Acknowledgements}}

I am most grateful to my friends and
collaborators Marc Baillargeon and Fawzi Boudjema  for all
the work on  electroweak physics issues.
I also thank   George Jikia for
kindly supplying the curves for $\gam\ra WWWW$.

\vspace{.5in}
%\noindent
%{\bf {\large References}}
%{\elevenbf \noindent References}
%\vglue 0.4cm


\begin{thebibliography}{99}
%\elevenrm\baselineskip=12pt
%
\bibitem{GinzburgNIM}
I.F. Ginzburg, {\it et al.}, {\em Nucl. Instrum. Methods} {\bf 205}
  (1983) 47.
%
\bibitem{Paris}
More details and references can be found in M.~Baillargeon, G.~B\'elanger and
F.~Boudjema,
{\it Proc. of Two-photon Physics from $DA\Phi NE$ to LEP200 and Beyond},
eds. F. Kapusta and J. Parisi, World Scientific, (1994) 267.
%
\bibitem{Telnovshef}
V. Telnov, these proceedings.
%
\bibitem{Chenshef}
P.Chen,  talk presented at the \gag workshop, Sheffield, April 7-8,
1995.
%
\bibitem{Jikiazz}
G.V. Jikia, \pl {\bf B298} (1993) 224; \np {\bf B405} (1993) 24.
\bibitem{DicusKao}
M.S.~Berger, \pr {\bf D48} (1993) 5121;
D.A.~Dicus and C.~Kao, \pr {\bf D49} (1994) 1265.
\bibitem{nousgg3v}
M.~Baillargeon and F.~Boudjema, \pl {\bf B317} (1993) 371.
\bibitem{Hawai}
F.~Boudjema, Proceedings of {\em Workshop on Physics and Experiments with
  Linear $e^+e^-$ Colliders}, eds. F.A.~Harris {\it et al.} (World Scientific,
  1994) 712.
\bibitem{steve}
G.~Couture, M.~Gintner, S.~Godfrey,  hep-ph/9505255.
\bibitem{barklow}
T.~Barklow, SLAC-PUB-6618, aug. 1994.
\bibitem{nousggwwz}
M.~Baillargeon, G.~B\'elanger, F.~Boudjema, G.~Couture,
in progress.
\bibitem{Dawsong5}
K.~Cheung, S.~Dawson, T.~Han and G.~Valencia, \pr {\bf D51} (1995) 5.
\bibitem{BMT}
M.~Bilenky {\it et al.}, \np {\bf 409} (1993) 22.
\bibitem{Dennerradcor}
Denner, Dittmaier, Schuster, BI-TP 95/04.
\bibitem{nousggvv}
G.~B\'elanger and F.~Boudjema, \pl {\bf B288} (1992) 210.
\bibitem{Ginzburgshef}
I. Ginzburg, these proceedings.
\bibitem{Borden}
 D.L.~Borden, Proceedings of {\em Workshop on Physics and Experiments with
  Linear $e^+e^-$ Colliders}, Eds. F.A.~Harris {\it et al.} (World Scientific,
  1994) 323.
\bibitem{Khozeshef}
V. Khoze, these proceedings; D.~L.~Borden {\it et al.}, hep-ph/9405401.
\bibitem{Halzen}
O.J.~P.~Eboli {\it et al.}, \pr {\bf D48}
  (1993) 1430.

\bibitem{Higgsreson}
M.~Baillargeon, G.~B\'elanger, F.~Boudjema, \pr {\bf D51} (1995) 4712.
\bibitem{Gunionparity}
J.F.~Gunion and J.~G.~Kelly, \pl {\bf B333} (1994) 110;
M.~Kr\"amer, J.~K\"uhn, M.~L.~Stong and P.~M.~Zerwas, \zp
{\bf C64} (1994) 21.
\bibitem{BergerBerkeley}
M.~Berger, M.~Chanowitz, {\em Nucl. Instrum. Methods} {\bf A355}
  (1995) 52.
\bibitem{Jikiawwww}
G.~V.~Jikia, {\em Nucl. Instrum. Methods} {\bf A355}
  (1995) 84.
\bibitem{Cheungwwww}
K.~Cheung, \pr {\bf D50} (1994) 4290.
\bibitem{GunionHaber}
J.~F.~Gunion and H.~Haber, \pr {\bf D48} (1993) 2907.
\bibitem{Gunionsusy}
J.~F.~Gunion, J.~G.~Kelly, J.~Ohnemus, \pr {\bf D51} (1995) 2101.
\bibitem{Murayama}
H.~Murayama, hep-ph/9410285.
\end{thebibliography}
\end{document}